\documentclass[pre,amsmath]{revtex4-1}
\usepackage[pdftex]{graphicx} 
\usepackage{color}
\usepackage{bm}
\usepackage{epsfig}
\usepackage{latexsym}
\usepackage{nameref,hyperref}
\begin{document}
\newcommand{\be}{\begin{equation}}
\newcommand{\ee}{\end{equation}}
\newcommand{\bea}{\begin{eqnarray}}
\newcommand{\eea}{\end{eqnarray}}
\newcommand{\RN}[1]{\textup{\uppercase\expandafter{\romannumeral#1}}}
\title{Effect of switching time scale of receptor activity on chemotactic performance of {\sl Escherichia coli}}
\author{Shobhan Dev Mandal and Sakuntala Chatterjee}
\affiliation{Department of Theoretical Sciences, S. N. Bose National Centre for Basic Sciences, Block JD, Sector 3, Salt Lake, Kolkata 700106, India.}
\begin{abstract}
In the chemotactic motion of {\sl Escherichia coli} the switching of transmembrane chemoreceptors between active and inactive states is one of the most important steps of the signaling pathway. We study the effect of this switching time-scale on the chemotactic performance of the cell. We quantify performance by the chemotactic drift velocity of the cell. Our extensive numerical simulations on a detailed theoretical model show that as the activity switching rate increases, the drift velocity increases and then saturates. Our data also show the mean duration of a downhill run decreases strongly with the switching rate, while that of an uphill run decreases relatively slowly. We explain this effect from temporal variation of activity along uphill and downhill trajectories. We show that for large and small switching rates the nature of activity variation show 
qualitatively different behaviors along a downhill run but similar behavior along an uphill run. This results in a stronger dependence of downhill run duration on the switching rate, and relatively milder dependence for uphill run duration.
\end{abstract}
\maketitle
\section{Introduction}

Chemotaxis refers to directed motion of an organism in response to a chemical gradient \cite{eisenbachbook}. Many prokaryotic organisms like {\sl {Escherichia coli, Salmonella typhimurium, Bacillus subtilis, Rhodobacter sphaeroides}} or {\sl{ Serratia marcescens}} \cite{salmonella, bsubtilis, rhotobacter, ariel2015swarming} and eukaryotic organisms like  {\sl{Chlamydomonas rheinhartii}} or {\sl{Tritrichomonas foetus}} \cite{chlam, foetus} use their run-and-tumble motility to perform chemotaxis. Among all these organisms the motion of {\sl Escherichia coli} or {\sl E.coli} is the most well-studied one \cite{berg2008coli}. In presence of a spatial concentration gradient of nutrient or some other attractant, an {\sl E.coli} cell shows a tendency to move up the gradient towards region of higher concentration \cite{adler1973method, adler1973chemotaxis, berg1972chemotaxis}. Each cell has about ten flagella associated with it via flagellar motors. When these motors rotate in the counter-clockwise (CCW) direction, the flagella form a helical bundle and propelled by this bundle the cell runs in one specific direction \cite{berg1972chemotaxis}. When few of the motors turn clockwise, the helical bundle is dispersed and the cell tumbles. The chemotaxis motion of {\sl E.coli} is mainly achieved by modulating the tumbling bias of the cell such that the runs up the gradient are extended and those down the gradient are shortened \cite{de2004chemotaxis, chatterjee2011chemotaxis}.

The intracellular signaling network of {\sl E.coli} consists of two principal modules: sensing and adaptation \cite{tu2013quantitative}. While the sensing module controls the response of the cell to the attractant signal, the adaptation module utilizes negative feedback mechanism to maintain the protein levels close to their average values \cite{barkai1997robustness}. These two modules are coupled via the activity of the transmembrane chemoreceptors. These receptors can switch between active and inactive states, depending on the local attractant concentration and the methylation levels of the receptors. Probability to find a receptor in active state, is defined as its activity. In an active state, the receptors promote auto-phosphorylation of the cytoplasmic protein CheA to form CheA-P, which in turn donates its phosphate group to CheY and CheB. In the phosphorylated state, CheY-P causes an increase in the tumbling bias \cite{eisenbach1996control}. On the other hand, the phosphorylated CheB-P demethylates the receptors which increases the probability of the receptors to switch to inactive state. Once in the inactive state, CheA phosphorylation drops, cutting down supply of phosphate group to CheY and CheB. In the absence of CheY-P tumbling bias decreases and the cell swims smoothly. Low abundance of CheB-P hinders demethylation and under the action of methylating enzyme CheR, the receptors' methylation levels go up which in turn increases the probability to switch the receptors to active state \cite{bren2000signals}. A high local concentration of attractant in the extra-cellular environment increases the probability of binding attractant molecules to the receptors and these binding events decrease the activity of the receptors. When the local concentration of attractant is low, there are only few nutrient molecules to bind to the receptors and hence the activity of the receptors remain high. This signaling pathway therefore ensures that the cell swims smoothly when the local attractant concentration is high and tumbles frequently when the attractant is present in low concentration.

One important aspect of the intracellular reaction network for {\sl E.coli} chemotaxis is the cooperativity among the chemoreceptors. There are few thousand receptors in an {\sl E.coli} cell and they are arranged in a hexagonal array \cite{briegel2012bacterial}. The neighboring receptors experience cooperative interaction and they tend to form `signaling teams' \cite{parkinson2015signaling}.  All receptors belonging to a particular team switch their activity states in unison. This results in amplification of the input signal coming from ligand binding \cite{sourjik2004functional,tu2013quantitative}. Larger the team size, higher is the amplification. This effect is responsible for the high sensitivity observed for {\sl E.coli} chemotaxis, where the cell is known to respond to even very weak concentration gradient of the attractant \cite{duke1999heightened,sourjik2004functional}. Recent experiments have shown that receptor cooperativity not only enhances the sensitivity, but also creates large activity fluctuations inside the cell \cite{colin2017multiple,keegstra2017phenotypic}. In other words, heightened sensitivity comes with a cost of intracellular noise. This results in an optimum size of the signaling team at which the chemotactic performance of the cell is at its best \cite{shobhan}. In our earlier work we explained this optimality as a result of competition between sensing and adaptation modules of the signaling network \cite{shobhan}. We have also shown that this sensing-adaptation interplay gives rise to highly non-trivial, interesting dynamics of the methylation and activity of the receptor clusters \cite{shobhanmethyl}.

The transition of a signaling team between the active and inactive states is captured well by MWC model \cite{monod1965nature} which has been quite successful so far in explaining a wide class of experimental data \cite{sourjik2004functional,keymer2006chemosensing,mello2005allosteric}. According to this model, the free energy difference between the active state and inactive state of a receptor depends logarithmically on the local attractant concentration and on the methylation level of the receptor. For a receptor cluster the free energy becomes sum of free energies of the individual receptors which form the cluster. This free energy is assumed to control the synchronous transition of all receptors in the cluster from one activity state to the other as per the principle of local detailed balance \cite{colin2017multiple}. At the biochemical level these transitions are associated with certain conformational changes of the receptor molecules \cite{parkinson2015signaling}.  In \cite{colin2017multiple} different values of this transition time-scale was considered in simulations and its effect on activity fluctuations was studied.  It was found that 
although the amplitude of fluctuation does not depend on the above time-scale, the the temporal fluctuation of activity increases as this time-scale becomes shorter \cite{colin2017multiple}.

In the present work, we investigate how different choices of the activity switching time-scale affect the chemotactic performance. We characterize the performance by chemotactic drift velocity, {\sl i.e.} the velocity with which the cell climbs up the attractant gradient. Clearly, a large value of this drift velocity indicates a better chemotactic performance. Our extensive numerical simulations on a detailed theoretical model show that the chemotactic drift velocity increases and then saturates as the switching of activity becomes faster. We explain this observation by detailed analysis of temporal variation of activity along the cell trajectory. We show that the downhill runs are strongly curtailed for faster activity switching and the nature of variation of activity along a downhill run becomes qualitatively different for fast and slow switches. On the other hand, uphill runs are less strongly affected by the switching timescale and this asymmetry results in an enhancement of chemotactic drift.

In the next section of this paper, we present our model. In Sec. \ref{sec:res} we present our results followed by conclusions. 

\begin{figure}
\includegraphics[scale=1.3]{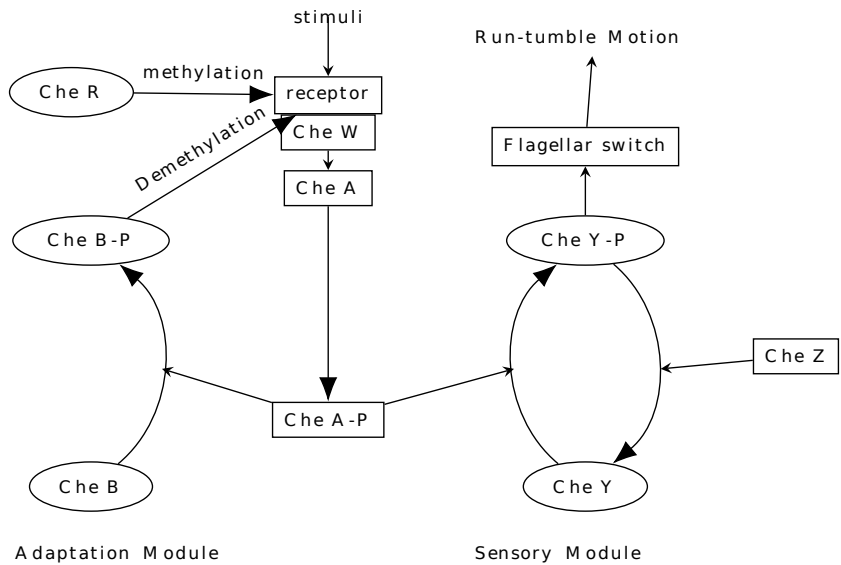}
\caption{Schematic diagram of E.coli signalling pathway.  }
\label{fig:sche}
\end{figure}

\section{Model description}
\label{sec:model}
Our model consists of three main parts: switching of activity of the chemoreceptors, (de)methylation of the receptors by enzymes, and run-tumble motility of the cell. In an {\sl E.coli} cell there are few thousand receptors which form dimers \cite{liu2012molecular, briegel2012bacterial}. Due to cooperative interaction, three dimers form trimer of dimers (TDs) and $n$ such TDs form a cluster or a signaling team. In our model, for simplicity, we assume all clusters have the same size. Activity state of a cluster is denoted as $\alpha$ which can take two values $0$ and $1$, according as the cluster is inactive or active. From Monod-Wyman-Changeux (MWC) model \cite{monod1965nature} the free energy difference between these two states can be written as  
\be 
F = 3n \left ( 1+ \log \frac{1+c(x)/K_{min}}{1+c(x)/K_{max}} \right ) - \sum_{j=1}^{3n} m_j   \label{eq:fnrg}
\ee
where $m_j$ is the methylation level of the $j$-th dimer in the cluster which can take any integer value between $0$ and $8$. At the cell position $x$ the concentration of the attractant is $c(x)$ . The two parameters $K_{min}$ and $K_{max}$ set the limit of sensitivity, such that for $c(x) \ll K_{min}$ or $c(x) \gg K_{max}$ the free energy becomes independent of $c(x)$ and the cell is not able to sense the concentration variation. The transition rate of the cluster from $\alpha=0$ state to $\alpha=1$ state is taken to be $\dfrac{w_\alpha}{1+\exp(F)}$ and the rate for the reverse transition is $\dfrac{w_{\alpha} \exp(F)}{1+\exp(F)}$ \cite{colin2017multiple}. Here, $w_\alpha$ has the dimension of rate which sets the overall timescale of activity switching process. We demonstrate in this work how different choices of $w_\alpha$ yield different behavior for the system.

The total number of molecules of methylating enzyme CheR and demethylating enzyme CheB are far too less (of the order of hundred) compared to the number of receptor dimers (few thousand) whose methylation levels they modify \cite{li2004cellular}. However, an {\sl E.coli} cell is known to show near-perfect adaptation \cite{berg1975transient, goy1977sensory}. To explain this few mechanisms like `assistance neighborhood' or 'brachiation' have been suggested and experimentally verified \cite{levin2002binding, endres2006precise, hansen2008chemotaxis, kim2002dynamic, li2005adaptational}. In an assistance neighborhood model, a single enzyme does not only modify the methylation level of the dimer it is attached to, but also of neighboring dimers. In a brachiation mechanism, an enzyme molecule can perform random walk on the receptor array and modify methylation levels of the dimers along its trajectory.

An unbound CheR molecule from the cell cytoplasm can bind to a receptor dimer if no other enzyme molecule is bound to it already. The binding can take place at the tether site or the modification site of the receptor dimer \cite{feng1999enhanced, wu1996receptor, pontius2013adaptation}. While both these bindings are slow, tether binding is relatively faster \cite{schulmeister2008protein, pontius2013adaptation} and hence we consider only tether binding in our model. The binding takes place with the rate $w_r$ and once bound, a CheR molecule raises the methylation level of the dimer by one unit provided methylation level of the dimer is less than $8$ and the dimer belongs to an inactive cluster. A bound CheR molecule can unbind from the dimer with rate $w_u$ and after that it can either rebind to another dimer within the same cluster, or it can return to the cell cytoplasm. The rebinding time-scale is much faster than $w_r$ and the possibility of rebinding within the same cluster allows multiple methylation reactions by a single enzyme molecule in that cluster before the enzyme molecule returns to the bulk. This is an effective way to capture the essence of both assistance neighborhood and brachiation mechanisms within our simple model. A CheB molecule in the cell cytoplasm can undergo phosphorylation with the rate $w_p$ by an active receptor, and dephosphorylation of CheB-P happens with the rate $w_{dp}$. An unbound CheB-P molecule can bind to a dimer and decrease its methylation level by one unit if the dimer is active and its methylation level is non-zero. The unbinding and rebinding process remains same as those for a CheR molecule.

The tumbling bias of the cell is controlled by the concentration of phosphorylated CheY molecules. When present in high concentration, CheY-P molecules bind to the flagellar motors and cause them to rotate in the clockwise direction which causes tumbles. Denoting $Y_P$ as the fraction of CheY molecules which are phosphorylated, the rate equation governing this fraction is given by \cite{jiang2010quantitative, flores2012signaling}
\begin{equation}  
\frac{dY_P}{dt}=K_Ya(1-Y_P)-K_ZY_P  \label{eq:yp}
\end{equation}
where $K_Y$ and $K_Z$ are rates of phosphorylation and dephosphorylation respectively and $a$ is the total activity of the cell, {\sl i.e.} average activity of all the signaling teams. In our simulation we directly use the steady state value $Y_P=\frac{a}{a+K_Z/K_Y}$. A cell in CCW rotational state or in run mode can switch to CW rotational state or tumble mode with rate $\omega \exp(-G)$ where $G=\Delta_1-\frac{\Delta_2}{1+Y_0/Y_P}$ and the opposite switch from tumble to run happens with a rate $\omega \exp(G)$ \cite{sneddon2012stochastic, dufour2014limits, micali2017drift}. $Y_{0}=0.34$ is the adapted value of $Y_P$ in our system. In Table \ref{table} we list all parameter values in our model.

We present here results for chemotaxis of a single cell in one dimension. We expect similar results in two dimensions as well \cite{shobhan, shobhanmethyl}. We consider a  one dimensional box of length $L=2000 \mu m$ with reflecting boundary conditions. We set up a linear nutrient profile of the form $c(x)=c_0(1+x/x_0)$, with $c_0=200$ $\mu M$ and $x_0=40000$ $\mu m$. This can be considered a weak gradient for which the steady state position distribution of the cell position is approximately linear. The smallest time-step  in our simulation is $dt = 0.01s$. During a run, the cell moves either leftward (downhill) or rightward (uphill) with speed $v=20 \mu m /s$ \cite{berg2008coli} such that in $dt$ time-step it travels a distance $vdt$. 
In the tumble mode, the cell position does not change with time. After each tumble, when the cell switches to run mode, the sign of $v$ is chosen at random. All our measurements are done in steady state, {\sl i.e.} after a long enough time, when the cell has been able to adapt to the attractant profile present in the medium.

\section{Results} \label{sec:res}
\subsection{Homogeneous attractant environment}

\begin{figure}
\includegraphics[scale=.6,angle=270]{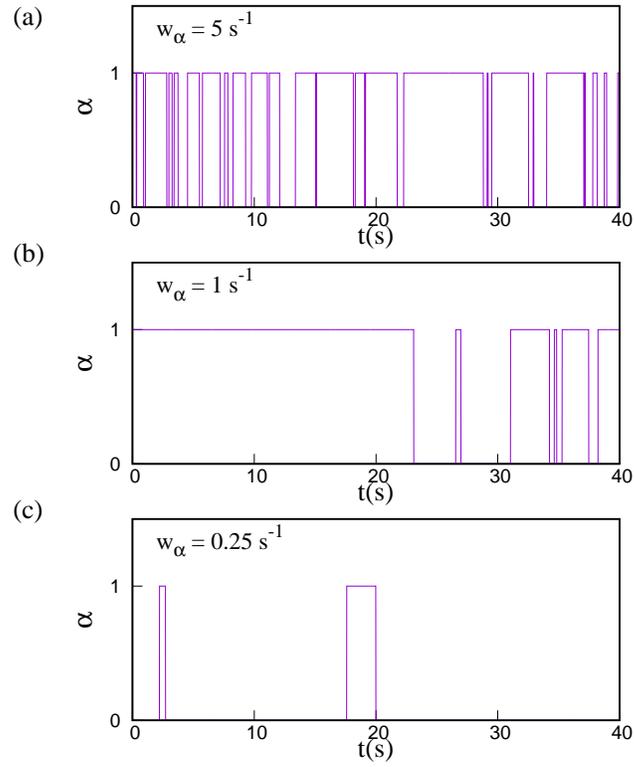}
\caption{Typical time series of activity of a single cluster recorded over a time-window ($40$ $s$) along the run-tumble trajectory of swimming cell in homogeneous attractant environment. As the switching rate ($w_{{\color{blue}\alpha}}$) increases, the switching between two activity states $0$ and $1$ happens more frequently. }
\label{fig:swi}
\end{figure}

\begin{figure}
\includegraphics[scale=0.5,angle=270]{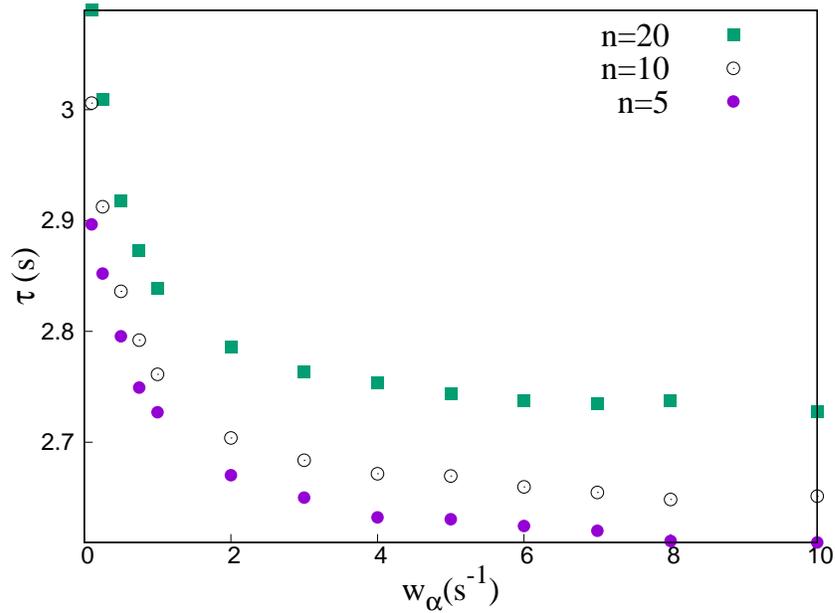}
\caption{ Mean run duration ($\tau$) of the chemotactic cell in homogeneous attractant environment as a function of switching rate ($w_{{\color{blue}\alpha}}$) for three different values of the cluster size $n$. The mean run duration  decreases with $w_{{\color{blue}\alpha}}$. Each data point has been averaged over at least $10^6$ histories. All simulation parameters are given in Sec. II. }
\label{fig:tau}
\end{figure}

To explicitly show the effect of activity switching rate on the temporal fluctuations of activity, we show in Fig. \ref{fig:swi} the time-series of the activity state of one particular cluster. As expected, the time interval between two successive switching events drops with increasing $w_{\alpha}$. For large $w_{\alpha}$ when the activity switching happens very frequently, the total number of active clusters fluctuates rapidly with time. The total activity that determines the tumbling bias of the cell, can go from high to low value in a short span of time. This means the residence time of the cell in run mode (or in tumble mode) decreases as $w_{\alpha}$ increases. In Fig. \ref{fig:tau} we verify this for mean duration of a run. Note that while $w_{\alpha}$ affects the temporal fluctuations of activity, the range of fluctuations does not show much dependence on $w_{\alpha}$. This range is mainly set by the value of $n$, the signaling team size \cite{shobhan, colin2017multiple}. Since the number of receptors in the cell is fixed, as the team size becomes large, there are fewer teams present. The total activity of the cell is then averaged over a fewer number of teams which increases the activity fluctuations. We have explicitly verified (data not shown here) that the activity distribution does not show any significant dependence on $w_{\alpha}$. These results are in line with \cite{colin2017multiple} where $w_\alpha$ was varied in simulation and it was found that the temporal fluctuation of activity increases with $w_\alpha$ but the amplitude of fluctuation remains unchanged.  

\subsection{Chemotactic performance gets better for fast activity switching}

\begin{figure}
\includegraphics[scale=1.3]{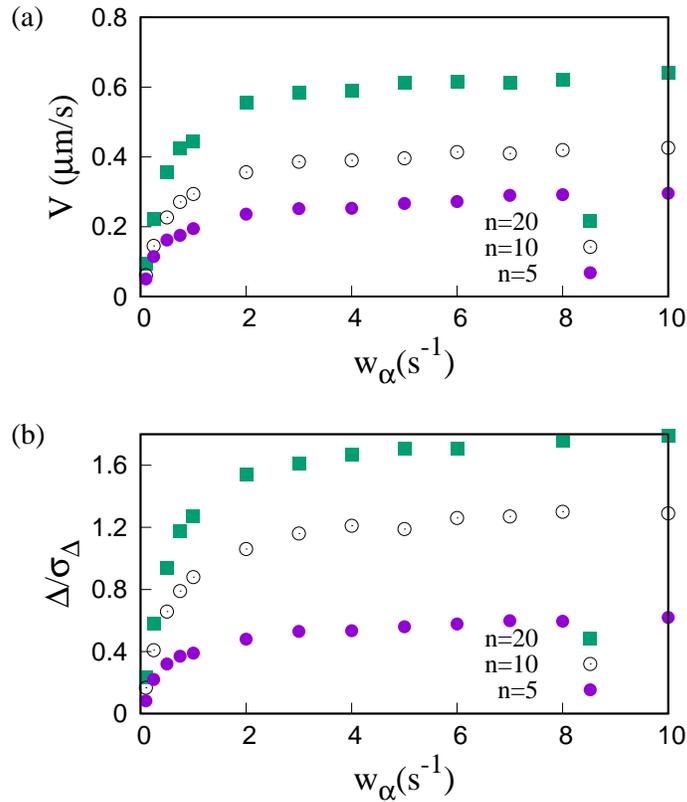}
\caption{Chemotactic efficiency of the cell as a function of switching rate $w_\alpha$. (a) Drift velocity ($V$) as function of switching  $w_{{\color{blue}\alpha}}$. Data for three different clustersize has been presented. $V$ increases with $w_\alpha$ and then saturates at large values of $w_\alpha$. (b) Ratio of net average displacement in a run($\Delta$) and fluctuation of this displacement($\sigma_\Delta$) as a function of switching rate $w_\alpha$. The ratio increases with $w_\alpha$ and then saturates at large values of $w_\alpha$. All simulation parameters are same as Fig. \ref{fig:tau}. These data points are averaged over at least $3\times 10^7$ histories.}
\label{fig:vwa}
\end{figure}

\begin{figure}
\includegraphics[scale=.6,angle=270]{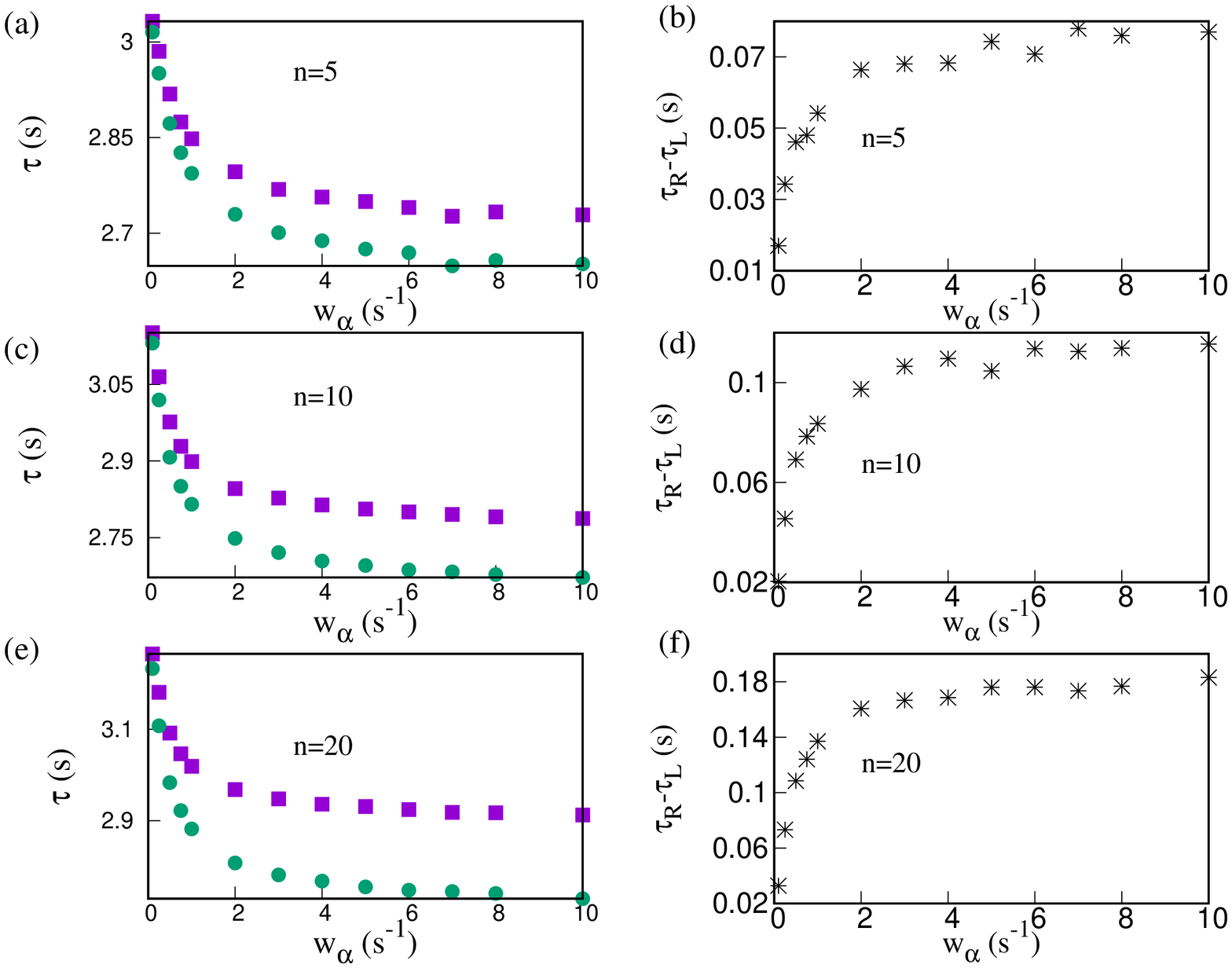}
\caption{Mean duration of an uphill run ($\tau_R$) and a downhill run ($\tau_L$) and their difference vs $w_{\alpha}$. Data for three different values of cluster sizes have been shown. The left panel shows that both $\tau_R$ (blue square) and $\tau_L$ (green circle)  decrease with $w_{\alpha}$ but $\tau_L$ falls at faster rate than $\tau_R$. The right panel shows the difference between $\tau_R$ and $\tau_L$ goes up with $w_{\alpha}$ and reaches a saturation. All simulation parameters are same as Fig. \ref{fig:tau}. These data points are averaged over at least $3 \times 10^7$ histories.}
\label{fig:rldwa}
\end{figure}

We quantify chemotactic performance by the drift velocity $V$ which measures how fast the cell climbs up the attractant gradient \cite{shobhan, dev2018optimal, chatterjee2011chemotaxis}. We define the drift velocity as $V=\dfrac{\Delta}{\tau}$, where $\Delta$ denotes the net displacement of the cell during a run and $\tau$ denotes average run duration. While a particular run can be directed both uphill and downhill, the form of the free energy gap between the two activity states of the cluster in Eq. \ref{eq:fnrg} is such that downhill (uphill) runs are associated with an increase (decrease) in activity, and therefore, an increase (decrease) in the tumbling bias. This means the uphill runs tend to get extended and downhill runs tend to get shortened, which results in a drift motion up the gradient. A large value of $V$ indicates faster motion of the cell up the gradient, which reflects a better chemotactic performance  \cite{shobhan, dev2018optimal, chatterjee2011chemotaxis}. To measure how reliably the cell is able to climb up the attractant gradient, we also calculate the ratio between the net directed displacement $\Delta$ and  its standard deviation $\sigma_\Delta$. A large value of this ratio would also imply a strong chemotactic performance. We are interested to find out how the chemotactic performance depends on the activity switching time-scale. In Fig. \ref{fig:vwa}a and \ref{fig:vwa}b we present our simulation results for $V$ and $\dfrac{\Delta}{\sigma_\Delta} $ for different values of $w_{\alpha}$ and three different choices for the signaling team size $n$. We find that both these quantities increase with $w_{\alpha}$ and then saturates. To explain this variation, consider the limit $w_{\alpha}=0$ which means activity of the receptors do not flip. This will naturally impair the chemotactic ability of the cell and $V$ will be zero. It is expected therefore for small $w_\alpha$, we should observe $V$ increasing with $w_{\alpha}$. However, when $w_{\alpha}$ is very large, the average run duration becomes small and then it is not obvious how $V$ is affected. To probe it further, we separately measure average duration of a rightward (uphill) run and a leftward (downhill) run as a function of $w_{\alpha}$. Our data in Fig. \ref{fig:rldwa} left panel show that both runs become shorter for large $w_{\alpha}$ as expected, but mean duration $\tau_L$ of leftward runs shows a faster drop with $w_{\alpha}$ than $\tau_R$ for rightward runs. This means the difference $\tau_R - \tau_L$ increases with $w_{\alpha}$ (see Fig. \ref{fig:rldwa} right panel). In other words, net displacement during a run increases, and mean run duration decreases with $w_{\alpha}$, as a consequence of which $V$ increases. In the next subsection, we present detailed explanation of this counter-intuitive behavior. 

\subsection{Explanation behind better performance at large $w_a$} 

\begin{figure}
\includegraphics[scale=0.35,angle=270]{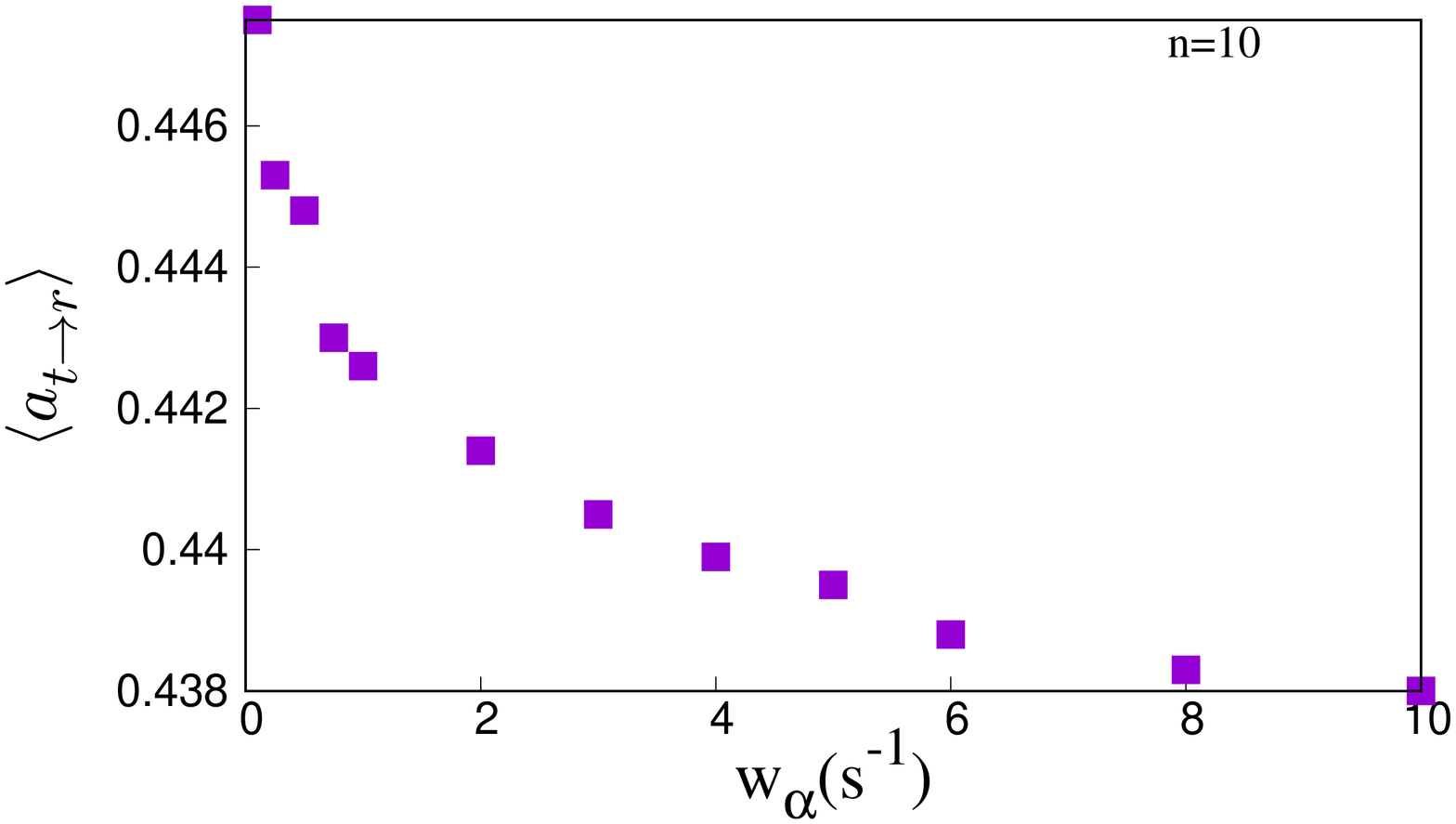}
\caption{Average of the activity $a_{t \to r}$ at the instant of a tumble to run switch as a function of $w_{\alpha}$. As $w_{\alpha}$ increases value of $\langle a_{t \to r} \rangle$  drops monotonically. Error bars of the data have magnitude of the order of $10^{-5}$ and hence they are actually smaller than the symbol size in the plot. All simulation parameters are same as Fig. \ref{fig:tau}. These data points are averaged over at least $5 \times 10^5$ histories.  }
\label{fig:ai}
\end{figure}

Large value of activity increases the tumbling bias of the cell. Therefore, when the cell enters the tumble mode, the number of active receptor clusters tends to be high. In the tumble mode the cell position does not change, which means ligand part of the free energy in Eq. \ref{eq:fnrg} remains fixed. The active receptors get demethylated and $F$ increases, which increases the probability of transition to inactive state. When $w_{\alpha}$ is high, more clusters flip from active to inactive state and the total activity drops quickly. Therefore, when the cell comes out of the tumble mode and starts a new run, its activity has a low value when $w_{\alpha}$ is large. Let $ a_{t \to r} $ denotes the activity when the tumble to run switch takes place.  In Fig. \ref{fig:ai} we plot the average of this quantity $\langle a_{t \to r} \rangle $ as a function of $w_{\alpha}$ and show that it decreases with $w_{\alpha}$. 

\begin{figure}
\includegraphics[scale=.7,angle=270]{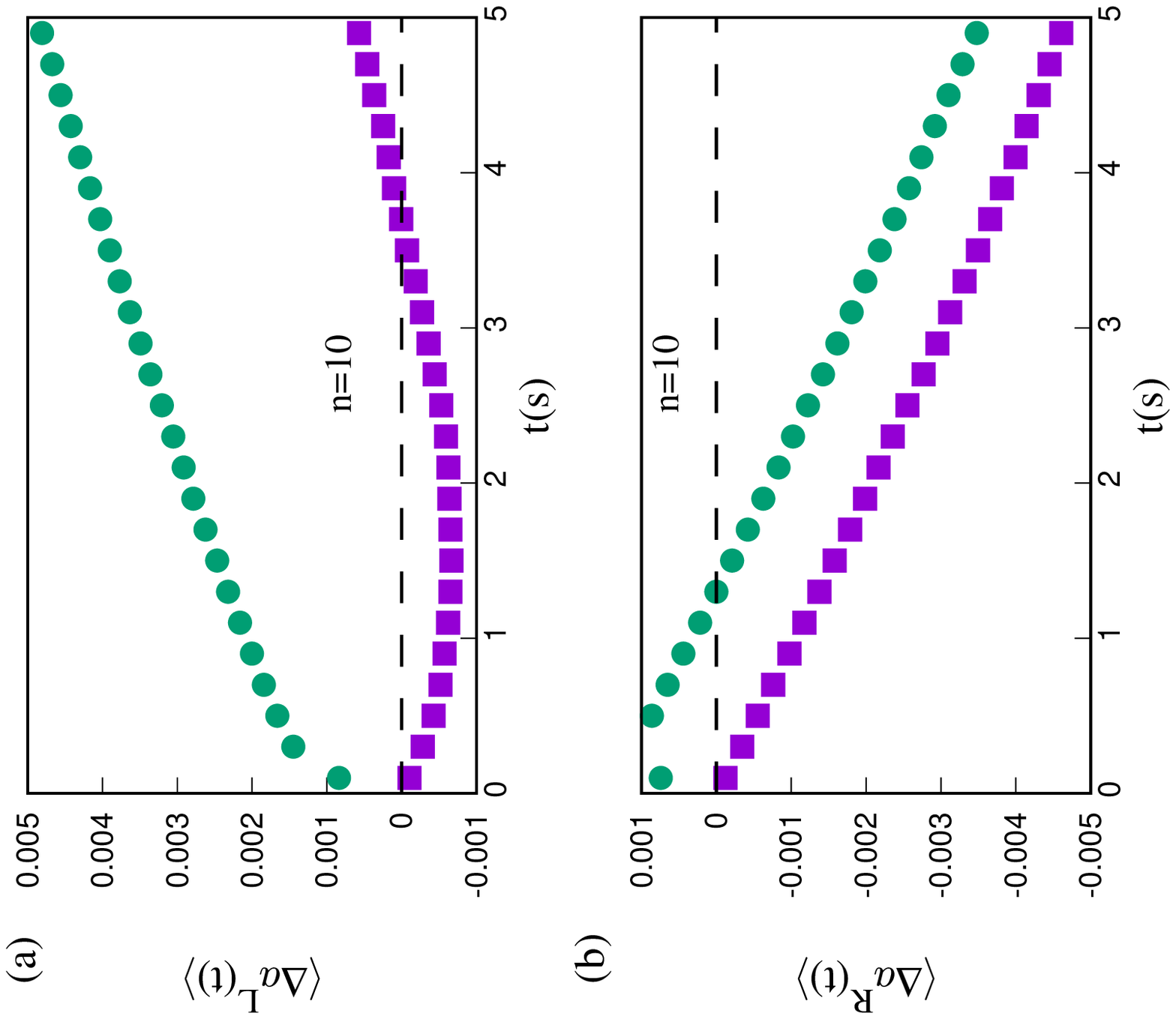}
\caption{Temporal variation of activity for $w_{\alpha}=0.75$ $s^{-1}$ (purple squares) and $w_{\alpha}=10$ $s^{-1}$ (green circles). (a) shows data for downhill runs For and (b) shows data for uphill runs. For downhill runs qualitative nature of variation is very different for the two $w_{\alpha}$ values, while for uphill runs the variation is qualitatively similar. All simulation parameters are same as Fig. \ref{fig:tau}. These data points are averaged over at least $4 \times 10^6$ histories.}
\label{fig:dela}
\end{figure}

The fact that new runs start with lower activity values as $w_\alpha$ increases, has an interesting implication for the activity variation during uphill and downhill runs, as shown below. Let $N^R(t)$ and $N^L(t)$ denote, respectively, the number of rightward (uphill) and leftward (downhill) runs whose duration is larger than $t$. Clearly, both these quantities decrease with $t$. Let $a^R_i(t)$ and $a^L_i(t)$ be the total activity of the cell at time $t$ during $i$-th uphill and downhill run, respectively. Now we can define the following quantities:
\begin{equation}
\langle{ \Delta a^R(t)} \rangle=\sum_{i=1}^{N^R(t)}\frac{a^R_i(t)-a^R_i(0)}{N^R(t)}
\end{equation}
and 
\begin{equation}
\langle \Delta a^L(t)\rangle=\sum_{i=1}^{N^L(t)}\frac{a^L_i(t)-a^L_i(0)}{N^L(t)}.
\end{equation}  
In Fig. \ref{fig:dela} we plot these quantities as a function of time for two different values of $w_{\alpha}$. Top panel (Fig. \ref{fig:dela}a) shows the variation of $\Delta a^L (t)$. For smaller $w_{\alpha}$ values (purple squares) when the runs start with relatively high value of activity, due to ongoing demethylation, activity drops at small $t$. Moreover, higher activity values make these runs prone to tumble and thus these trajectories drop out of $N^L(t)$ population quickly, which brings down the average activity. As a result, $\Delta a^L (t)$ takes negative values for small $t$. As $t$ increases, however, drop in ligand concentration along the downhill run tends to increase the activity again. Due to these two opposing effects, $\Delta a^L (t)$ after reaching a minimum starts slowly increasing again. Our data show that till moderate or large $t$, $\Delta a^L (t)$ remains negative with its magnitude decreasing with $t$ {\sl i.e.} activity still decreases on an average during the run, but the magnitude of the change becomes less. At very large $t$ change in activity becomes positive but still has a small magnitude. The situation is quite different for large $w_a$ (green circles). Here, initial activity at $t=0$ is low which prompts methylation and decreases $F$ (Eq. \ref{eq:fnrg}). Also, decreasing ligand concentration along the trajectory causes $F$ to decrease. Large $w_{\alpha}$ ensures that activity responds quickly to this free energy change and more and more clusters flip from inactive to active state. $\Delta a^L(t)$ is positive in this case and grows rapidly with time. The behavior of $\Delta a^L(t)$ is thus completely different for small and large $w_\alpha$. From Fig. \ref{fig:dela}a it is clear that the values of $\Delta a^L(t)$ for small and large $w_{\alpha}$ are widely apart for the range of time we have observed.

On the other hand, Fig. \ref{fig:dela}b shows that the qualitative behavior of $\Delta a^R(t))$ is quite similar for large and small $w_{\alpha}$. For small $w_{\alpha}$ (purple squares) runs start with larger activity and subsequent demethylation together with increasing ligand concentration along the trajectory cause the activity to decrease making $\Delta a^R(t))$ negative with its magnitude increasing steadily with time. For large $w_{\alpha}$ (green circles) initial activity being small, receptors tend to methylate which causes activity to increase sharply for small $t$. But for moderate or large times, $c(x)$ variation takes over and causes activity to decrease with time. Thus for both small and large $w_\alpha$ we find $\Delta a^R(t))$ shows similar temporal variation, except for very small times. For the entire time-range we have observed, the values of $\Delta a^R(t))$ for small and large $w_{\alpha}$ are relatively closer compared to Fig. \ref{fig:dela}a. In other words, average change in activity for small and large $w_{\alpha}$ are very different along a downhill run, but the difference is significantly less along an uphill run. This explains why $\tau_L$ decreases more strongly with $w_{\alpha}$ compared to $\tau_R$ (Fig. \ref{fig:rldwa}).

\section{Conclusions}
\label{sec:con}

In this paper we have investigated the effect of activity switching timescale on the chemotactic performance of a single {\sl E.coli} cell. We find that the performance improves as the activity switches faster. We quantify performance by (a) the chemotactic drift velocity that measures how rapidly the cell manages to climb up the chemical concentration gradient, and (b) by the reliability of its net uphill drift. In both cases we find better performance for faster activity switches. More specifically, the average duration of a downhill run decreases rapidly with the activity switching rate $w_{\alpha}$, but that of an uphill run decreases comparatively slowly. This asymmetry enhances the chemotactic drift. To explain this observation we monitor temporal variation of activity during uphill and downhill runs. Our detailed measurements show that during downhill runs, nature of activity variation shows widely different behavior for small and large values of $w_{\alpha}$, while the behavior is not so different during uphill runs. This results in stronger dependence on $w_{\alpha}$ of downhill run duration which explains the above effect.

It should be possible to experimentally verify some of our main results. Direct measurement of receptor activity has already been possible using FRET  based experiments \cite{sourjik2002receptor, sourjik2004functional, colin2017multiple}. Although most experiments focus on tethered cells, and our studies  consider a swimming cell, it is possible to verify our conclusions even for a tethered cell by appropriately engineered attractant environment. By identifying the counter clockwise rotation of the flagellar motors as the run mode and the clockwise rotation as the tumble mode, one can ramp up (down) the attractant concentration to mimic an uphill (downhill) run and hold the attractant level steady for a tumble. Using this set-up the activity variation of the cell can be measured with time and our results on $\Delta a^R(t)$ and $\Delta a^L(t)$ can be verified. The switching of a receptor between active and inactive states is a result of a conformational change of the receptor molecule \cite{parkinson2015signaling}. An experimental control over this conformational change timescale may provide some insights into its effect on the chemotactic performance of the cell. Moreover, our study can be used to estimate the activity switching rate for wild-type {\sl E.coli}, which has not been measured so far to the best of our knowledge. The range of values of drift velocity measured in our simulations agree with experimental data for $V$ recorded under similar attractant environment \cite{ahmed2008experimental, micali2017drift}. A comparison between the experiments and our simulations yields $w_\alpha$ ranging between $0.5 s^{-1}$ and $1s^{-1}$. A direct experimental measurement of $w_\alpha$ can help to verify this prediction. 

\section{Acknowledgements}
SDM acknowledges a research fellowship [Grant No. 09/575(0122)/2019-EMR -I]  from the Council of Scientific and Industrial Research (CSIR), India. SC acknowledges financial support from the Science and Engineering Research Board, India (Grant No: MTR/2019/000946).

\appendix
\newpage

\newpage
\section{Model parameters}
\label{app:tab}
\begin{center}
\begin{table}[h]
\caption{Parameter values used in model and simulations}
\begin{tabular}{|l|l|l|l|}
\hline
\textbf{Symbol} & \hspace{30mm} \textbf{Description} & \textbf{Value} & \textbf{References} \\
\hline
$N_{dim}$ & Total number of receptor dimers & $7200$ & \cite{pontius2013adaptation,li2004cellular} \\
\hline
$N_R$  & Total number of CheR  protein molecules &  $140$ & \cite{pontius2013adaptation,li2004cellular}\\
\hline
$N_B$  & Total number of CheB  protein molecules &  $240$ & \cite{pontius2013adaptation,li2004cellular} \\
\hline
$K_{min}$ & Minimum concentration receptor can sense &  $18$ $\mu M$ & \cite{jiang2010quantitative}, \cite{flores2012signaling} \\
\hline
$K_{max}$ & Maximum concentration receptor can sense &  $3000$ $\mu M$ & \cite{flores2012signaling,jiang2010quantitative} \\
\hline
$\omega$  & Switching frequency of motor &  $1.3$ $s^{-1}$ & \cite{sneddon2012stochastic} \\
\hline
 $\Delta_{1}$  & Non-dimensional constant regulating motor switching  &  $10$ & \cite{sneddon2012stochastic} \\
\hline
$\Delta_{2}$  & Non-dimensional constant regulating motor switching &  $20$ & \cite{sneddon2012stochastic}  \\
\hline
$Y_{0}$ & Adapted value of the fraction of CheY-P protein &  $0.34$ & \cite{sneddon2012stochastic} \\
\hline 
$K_{Y}$ & Phosphorylation rate of CheY molecule &  $1.7$ $s^{-1}$ & \cite{flores2012signaling,tu2008modeling} \\
\hline
$K_{Z}$ & Dephosphorylation rate of CheY molecule &  $2$ $s^{-1}$ & \cite{flores2012signaling,tu2008modeling}\\
\hline
$w_{r}$ & Binding rate of bulk CheR to tether site of an unoccupied dimer &  $0.068$ $s^{-1}$ & \cite{pontius2013adaptation,schulmeister2008protein} \\
\hline
$w_{b}$ & Binding rate of bulk CheB-P to tether site of an unoccupied dimer &  $0.061$ $s^{-1}$ & \cite{pontius2013adaptation, schulmeister2008protein} \\
\hline
$w_{u}$ & Unbinding rate of bound CheR and CheB-P &  $5$ $s^{-1}$ & \cite{pontius2013adaptation,schulmeister2008protein} \\	
\hline	
$k_{r}$ & Methylation rate of bound CheR & $2.7$ $s^{-1}$ & \cite{pontius2013adaptation,schulmeister2008protein}\\ 	
\hline
$k_{b}$ & Demethylation rate of bound CheB-P &  $3$ $s^{-1}$ & \cite{pontius2013adaptation,schulmeister2008protein}\\
\hline	
$w_{p}$ & CheB phosphorylation rate &  $3$ $s^{-1}$ & \cite{pontius2013adaptation,stewart2000rapid} \\
\hline
$w_{dp}$ & CheB-P dephosphorylation rate &  $0.37$ $s^{-1}$ & \cite{pontius2013adaptation}\\	
\hline	

\end{tabular}
\label{table}
\end{table}
\end{center}

\bibliographystyle{unsrt}
\bibliography{papbib}

\begin{thebibliography}{10}

\bibitem{eisenbachbook}
Michael Eisenbach.
\newblock {\em Chemotaxis}.
\newblock Imperial College Press and World Scientific Publishing Co., 2004.

\bibitem{salmonella}
RICHARD~J Galloway and BARRY~L Taylor.
\newblock Histidine starvation and adenosine 5'-triphosphate depletion in
  chemotaxis of salmonella typhimurium.
\newblock {\em Journal of bacteriology}, 144(3):1068--1075, 1980.

\bibitem{bsubtilis}
Marina Sidortsov, Yakov Morgenstern, and Avraham Be'er.
\newblock Role of tumbling in bacterial swarming.
\newblock {\em Physical Review E}, 96(2):022407, 2017.

\bibitem{rhotobacter}
Gabriel Rosser, Ruth~E Baker, Judith~P Armitage, and Alexander~G Fletcher.
\newblock Modelling and analysis of bacterial tracks suggest an active
  reorientation mechanism in rhodobacter sphaeroides.
\newblock {\em Journal of The Royal Society Interface}, 11(97):20140320, 2014.

\bibitem{ariel2015swarming}
Gil Ariel, Amit Rabani, Sivan Benisty, Jonathan~D Partridge, Rasika~M Harshey,
  and Avraham Be'Er.
\newblock Swarming bacteria migrate by l{\'e}vy walk.
\newblock {\em Nature communications}, 6(1):1--6, 2015.

\bibitem{chlam}
Marco Polin, Idan Tuval, Knut Drescher, Jerry~P Gollub, and Raymond~E
  Goldstein.
\newblock Chlamydomonas swims with two “gears” in a eukaryotic version of
  run-and-tumble locomotion.
\newblock {\em Science}, 325(5939):487--490, 2009.

\bibitem{foetus}
Scott~C Lenaghan, Stefan Nwandu-Vincent, Benjamin~E Reese, and Mingjun Zhang.
\newblock Unlocking the secrets of multi-flagellated propulsion: drawing
  insights from tritrichomonas foetus.
\newblock {\em Journal of The Royal Society Interface}, 11(93):20131149, 2014.

\bibitem{berg2008coli}
Howard~C Berg.
\newblock {\em E. coli in Motion}.
\newblock Springer Science \& Business Media, 2008.

\bibitem{adler1973method}
Julius Adler.
\newblock A method for measuring chemotaxis and use of the method to determine
  optimum conditions for chemotaxis by escherichia coli.
\newblock {\em Microbiology}, 74(1):77--91, 1973.

\bibitem{adler1973chemotaxis}
Julius Adler, Gerald~L Hazelbauer, and MM~Dahl.
\newblock Chemotaxis toward sugars in escherichia coli.
\newblock {\em Journal of bacteriology}, 115(3):824--847, 1973.

\bibitem{berg1972chemotaxis}
Howard~C Berg and Douglas~A Brown.
\newblock Chemotaxis in escherichia coli analysed by three-dimensional
  tracking.
\newblock {\em Nature}, 239(5374):500--504, 1972.

\bibitem{de2004chemotaxis}
P-G De~Gennes.
\newblock Chemotaxis: the role of internal delays.
\newblock {\em European Biophysics Journal}, 33(8):691--693, 2004.

\bibitem{chatterjee2011chemotaxis}
Sakuntala Chatterjee, Rava~Azeredo da~Silveira, and Yariv Kafri.
\newblock Chemotaxis when bacteria remember: drift versus diffusion.
\newblock {\em PLoS computational biology}, 7(12), 2011.

\bibitem{tu2013quantitative}
Yuhai Tu.
\newblock Quantitative modeling of bacterial chemotaxis: signal amplification
  and accurate adaptation.
\newblock {\em Annual review of biophysics}, 42:337--359, 2013.

\bibitem{barkai1997robustness}
Naama Barkai and Stan Leibler.
\newblock Robustness in simple biochemical networks.
\newblock {\em Nature}, 387(6636):913--917, 1997.

\bibitem{eisenbach1996control}
Michael Eisenbach.
\newblock Control of bacterial chemotaxis.
\newblock {\em Molecular microbiology}, 20(5):903--910, 1996.

\bibitem{bren2000signals}
Anat Bren and Michael Eisenbach.
\newblock How signals are heard during bacterial chemotaxis: protein-protein
  interactions in sensory signal propagation.
\newblock {\em Journal of bacteriology}, 182(24):6865--6873, 2000.

\bibitem{briegel2012bacterial}
Ariane Briegel, Xiaoxiao Li, Alexandrine~M Bilwes, Kelly~T Hughes, Grant~J
  Jensen, and Brian~R Crane.
\newblock Bacterial chemoreceptor arrays are hexagonally packed trimers of
  receptor dimers networked by rings of kinase and coupling proteins.
\newblock {\em Proceedings of the National Academy of Sciences},
  109(10):3766--3771, 2012.

\bibitem{parkinson2015signaling}
John~S Parkinson, Gerald~L Hazelbauer, and Joseph~J Falke.
\newblock Signaling and sensory adaptation in escherichia coli chemoreceptors:
  2015 update.
\newblock {\em Trends in microbiology}, 23(5):257--266, 2015.

\bibitem{sourjik2004functional}
Victor Sourjik and Howard~C Berg.
\newblock Functional interactions between receptors in bacterial chemotaxis.
\newblock {\em Nature}, 428(6981):437--441, 2004.

\bibitem{duke1999heightened}
TAJ Duke and Dennis Bray.
\newblock Heightened sensitivity of a lattice of membrane receptors.
\newblock {\em Proceedings of the National Academy of Sciences},
  96(18):10104--10108, 1999.

\bibitem{colin2017multiple}
Remy Colin, Christelle Rosazza, Ady Vaknin, and Victor Sourjik.
\newblock Multiple sources of slow activity fluctuations in a bacterial
  chemosensory network.
\newblock {\em Elife}, 6:e26796, 2017.

\bibitem{keegstra2017phenotypic}
Johannes~M Keegstra, Keita Kamino, Fran{\c{c}}ois Anquez, Milena~D Lazova,
  Thierry Emonet, and Thomas~S Shimizu.
\newblock Phenotypic diversity and temporal variability in a bacterial
  signaling network revealed by single-cell fret.
\newblock {\em Elife}, 6:e27455, 2017.

\bibitem{shobhan}
Shobhan~Dev Mandal and Sakuntala Chatterjee.
\newblock Effect of receptor clustering on chemotactic performance of e. coli:
  Sensing versus adaptation.
\newblock {\em Phys. Rev. E}, 103:L030401, Mar 2021.

\bibitem{shobhanmethyl}
Shobhan~Dev Mandal and Sakuntala Chatterjee.
\newblock Effect of receptor cooperativity on methylation dynamics in bacterial
  chemotaxis with weak and strong gradient.
\newblock {\em Submitted}.

\bibitem{monod1965nature}
Jacque Monod, Jeffries Wyman, and Jean-Pierre Changeux.
\newblock On the nature of allosteric transitions: a plausible model.
\newblock {\em J Mol Biol}, 12(1):88--118, 1965.

\bibitem{keymer2006chemosensing}
Juan~E Keymer, Robert~G Endres, Monica Skoge, Yigal Meir, and Ned~S Wingreen.
\newblock Chemosensing in escherichia coli: two regimes of two-state receptors.
\newblock {\em Proceedings of the National Academy of Sciences},
  103(6):1786--1791, 2006.

\bibitem{mello2005allosteric}
Bernardo~A Mello and Yuhai Tu.
\newblock An allosteric model for heterogeneous receptor complexes:
  understanding bacterial chemotaxis responses to multiple stimuli.
\newblock {\em Proceedings of the National Academy of Sciences},
  102(48):17354--17359, 2005.

\bibitem{liu2012molecular}
Jun Liu, Bo~Hu, Dustin~R Morado, Sneha Jani, Michael~D Manson, and William
  Margolin.
\newblock Molecular architecture of chemoreceptor arrays revealed by
  cryoelectron tomography of escherichia coli minicells.
\newblock {\em Proceedings of National Academy of Sciences},
  109(23):E1481--E1488, 2012.

\bibitem{li2004cellular}
Mingshan Li and Gerald~L Hazelbauer.
\newblock Cellular stoichiometry of the components of the chemotaxis signaling
  complex.
\newblock {\em Journal of bacteriology}, 186(12):3687--3694, 2004.

\bibitem{berg1975transient}
Howard~C Berg and PM~Tedesco.
\newblock Transient response to chemotactic stimuli in escherichia coli.
\newblock {\em Proceedings of the National Academy of Sciences},
  72(8):3235--3239, 1975.

\bibitem{goy1977sensory}
Michael~F Goy, Martin~S Springer, and Julius Adler.
\newblock Sensory transduction in escherichia coli: role of a protein
  methylation reaction in sensory adaptation.
\newblock {\em Proceedings of the National Academy of Sciences},
  74(11):4964--4968, 1977.

\bibitem{levin2002binding}
Matthew~D Levin, Thomas~S Shimizu, and Dennis Bray.
\newblock Binding and diffusion of cher molecules within a cluster of membrane
  receptors.
\newblock {\em Biophysical journal}, 82(4):1809--1817, 2002.

\bibitem{endres2006precise}
Robert~G Endres and Ned~S Wingreen.
\newblock Precise adaptation in bacterial chemotaxis through “assistance
  neighborhoods”.
\newblock {\em Proceedings of the National Academy of Sciences},
  103(35):13040--13044, 2006.

\bibitem{hansen2008chemotaxis}
Clinton~H Hansen, Robert~G Endres, and Ned~S Wingreen.
\newblock Chemotaxis in escherichia coli: a molecular model for robust precise
  adaptation.
\newblock {\em PLoS Comput Biol}, 4(1):e1, 2008.

\bibitem{kim2002dynamic}
Sung-Hou Kim, Weiru Wang, and Kyeong~Kyu Kim.
\newblock Dynamic and clustering model of bacterial chemotaxis receptors:
  structural basis for signaling and high sensitivity.
\newblock {\em Proceedings of the National Academy of Sciences},
  99(18):11611--11615, 2002.

\bibitem{li2005adaptational}
Mingshan Li and Gerald~L Hazelbauer.
\newblock Adaptational assistance in clusters of bacterial chemoreceptors.
\newblock {\em Molecular microbiology}, 56(6):1617--1626, 2005.

\bibitem{feng1999enhanced}
Xiuhong Feng, Angela~A Lilly, and Gerald~L Hazelbauer.
\newblock Enhanced function conferred on low-abundance chemoreceptor trg by a
  methyltransferase-docking site.
\newblock {\em Journal of bacteriology}, 181(10):3164--3171, 1999.

\bibitem{wu1996receptor}
Jiongru Wu, Jiayin Li, Guoyong Li, David~G Long, and Robert~M Weis.
\newblock The receptor binding site for the methyltransferase of bacterial
  chemotaxis is distinct from the sites of methylation.
\newblock {\em Biochemistry}, 35(15):4984--4993, 1996.

\bibitem{pontius2013adaptation}
William Pontius, Michael~W Sneddon, and Thierry Emonet.
\newblock Adaptation dynamics in densely clustered chemoreceptors.
\newblock {\em PLoS computational biology}, 9(9), 2013.

\bibitem{schulmeister2008protein}
Sonja Schulmeister, Michaela Ruttorf, Sebastian Thiem, David Kentner, Dirk
  Lebiedz, and Victor Sourjik.
\newblock Protein exchange dynamics at chemoreceptor clusters in escherichia
  coli.
\newblock {\em Proceedings of the National Academy of Sciences},
  105(17):6403--6408, 2008.

\bibitem{jiang2010quantitative}
Lili Jiang, Qi~Ouyang, and Yuhai Tu.
\newblock Quantitative modeling of escherichia coli chemotactic motion in
  environments varying in space and time.
\newblock {\em PLoS Comput Biol}, 6(4):e1000735, 2010.

\bibitem{flores2012signaling}
Marlo Flores, Thomas~S Shimizu, Pieter~Rein ten Wolde, and Filipe Tostevin.
\newblock Signaling noise enhances chemotactic drift of e. coli.
\newblock {\em Physical review letters}, 109(14):148101, 2012.

\bibitem{sneddon2012stochastic}
Michael~W Sneddon, William Pontius, and Thierry Emonet.
\newblock Stochastic coordination of multiple actuators reduces latency and
  improves chemotactic response in bacteria.
\newblock {\em Proceedings of the National Academy of Sciences},
  109(3):805--810, 2012.

\bibitem{dufour2014limits}
Yann~S Dufour, Xiongfei Fu, Luis Hernandez-Nunez, and Thierry Emonet.
\newblock Limits of feedback control in bacterial chemotaxis.
\newblock {\em PLoS Comput Biol}, 10(6):e1003694, 2014.

\bibitem{micali2017drift}
Gabriele Micali, R{\'e}my Colin, Victor Sourjik, and Robert~G Endres.
\newblock Drift and behavior of e. coli cells.
\newblock {\em Biophysical journal}, 113(11):2321--2325, 2017.

\bibitem{dev2018optimal}
Subrata Dev and Sakuntala Chatterjee.
\newblock Optimal methylation noise for best chemotactic performance of e.
  coli.
\newblock {\em Physical Review E}, 97(3):032420, 2018.

\bibitem{sourjik2002receptor}
Victor Sourjik and Howard~C Berg.
\newblock Receptor sensitivity in bacterial chemotaxis.
\newblock {\em Proceedings of the National Academy of Sciences},
  99(1):123--127, 2002.

\bibitem{ahmed2008experimental}
Tanvir Ahmed and Roman Stocker.
\newblock Experimental verification of the behavioral foundation of bacterial
  transport parameters using microfluidics.
\newblock {\em Biophysical journal}, 95(9):4481--4493, 2008.

\bibitem{tu2008modeling}
Yuhai Tu, Thomas~S Shimizu, and Howard~C Berg.
\newblock Modeling the chemotactic response of escherichia coli to time-varying
  stimuli.
\newblock {\em Proceedings of the National Academy of Sciences},
  105(39):14855--14860, 2008.

\bibitem{stewart2000rapid}
Richard~C Stewart, Knut Jahreis, and John~S Parkinson.
\newblock Rapid phosphotransfer to chey from a chea protein lacking the
  chey-binding domain.
\newblock {\em Biochemistry}, 39(43):13157--13165, 2000.

\end{thebibliography}
\end{document}